# Network resilience in the aging brain

Tao Liu[a]*[#], Shu Guo[b]*[#], Hao Liu[a][#], Rui Kang[b], Mingyang Bai[b], Jiyang Jiang[c], Wei Wen[c,d], Xing Pan[b], Jun Tai[e], Jianxin Li[f], Jian Cheng[f], Jing Jing[g], Zhenzhou Wu[h], Haijun Niu[a], Haogang Zhu[f], Zixiao Li[i], Yongjun Wang[i], Henry Brodaty[c,j], Perminder Sachdev[c,d], Daqing Li[b,k]*

[a]Beijing Advanced Innovation Center for Biomedical Engineering, School of Biological Science and Medical Engineering, Beihang University, China
[b]School of Reliability and Systems Engineering, Beihang University, China
[c]Centre for Healthy Brain Ageing (CHeBA), School of Psychiatry, UNSW Sydney NSW 2052, Australia
[d]Neuropsychiatric Institute, Prince of Wales Hospital, Sydney, NSW, Australia
[e]Department of Otorhinolaryngology, Children's Hospital, Capital Institute of Pediatrics, No. 2 Yabao Road, Chaoyang District, 100020 Beijing, China.
[f]School of Computer Science and Engineering, Beihang University, Beijing, China
[g]Neuroimaging Center of Excellence, China National Clinical Research Center for Neurological Diseases, Beijing, China
[h]BioMind Technology AI Center, China National Clinical Research Center for Neurological Diseases, Beijng, China
[i]Vascular Neurology, Department of Neurology, Beijing TianTan Hospital, Capital Medical University, China
[j]Dementia Centre for Research Collaboration, School of Psychiatry, UNSW Sydney NSW 2052, Australia
[k]College of Safety Science and Engineering, Civil Aviation University of China, Tianjin 300300, China

[#]co-first authors     *Correspondence

**Acknowledgements**
This research received support from the National Natural Science Foundation of China (Grant No. 81871434, 71822101 and 61971017), Beijing Natural Science Foundation (Grant No. Z200016), Public service development and reform pilot project of Beijing Medical Research Institute (BMR2019-11), National natural science foundation of China (81970900) and Beijing Social Science Foundation Project (19GLB033). MAS (The Sydney Memory and Ageing Study) cohort was supported by National Health and Medical Research Council (NHMRC) Australia Project Program Grants ID350833, ID568969 and ID1093083. We thank the participants and their informants for their time and generosity in contributing to this research. We also acknowledge the MAS research team and professor Shlomo Havlin in Bar-ilan University for helpful discussion.

**Abstract**

Degeneration and adaptation are two competing sides of the same coin called resilience in the progressive processes of brain aging or diseases. Degeneration accumulates during brain aging and other cerebral activities, causing structural atrophy and dysfunction. At the same time, adaptation allows brain network reorganize to compensate for structural loss to maintain cognition function. Although hidden resilience mechanism is critical and fundamental to uncover the brain aging law, due to the lack of datasets and appropriate methodology, it remains essentially unknown how these two processes interact dynamically across brain networks. To quantitatively investigate this complex process, we analyze aging brains based on 6-year follow-up multimodal neuroimaging database from 63 persons. We reveal the critical mechanism of network resilience that various perturbation may cause fast brain structural atrophy, and then brain can reorganize its functional layout to lower its operational efficiency, which helps to slow down the structural atrophy and finally recover its functional efficiency equilibrium. This empirical finding could be explained by our theoretical model, suggesting one universal resilience dynamical function. This resilience is achieved in the brain functional network with evolving percolation and rich-club features. Our findings can help to understand the brain aging process and design possible mitigation methods to adjust interaction between degeneration and adaptation from resilience viewpoint.

**Introduction**

The operation of the brain relies on dynamic adaptation to achieve a trade-off between wiring cost and effective function (Bullmore and Sporns 2012). Aging poses a degeneration challenge to this adaptation by structural atrophy and dysfunction, which makes the brain vulnerable to age-related dysfunction (Fjell et al. 2014). Prior studies have provided some evidence that the brain compensates for aging-related deficits by recruiting suboptimal regions or reorganizing new activity patterns (Cabeza et al. 2018; Park and Reuter-Lorenz 2009). These two processes of degeneration and adaptation are not independent. The vulnerability to neurofibrillary tangles, a hallmark of Alzheimer's disease (AD), the most common age-related neurodegenerative disease, is significantly higher in limbic areas at the very early stage (Mesulam 1999). As the limbic areas have higher neuroplasticity than other regions, these findings indicate that those brain regions with high adaptation throughout the life cycle are also accompanied by a high risk of vulnerability (Mesulam 1999; Neill 2012; Rapoport and Nelson 2011). In this degeneration-adaptation antagonism interaction, the trajectory of brain aging is not linear decline, but more complex resilience process from the adversarial interaction between decline caused by vulnerability, and recovery resulting from adaptation (Damoiseaux 2017; Fjell et al. 2013). This dynamic process can be regarded as the maintenance of brain through preservation of neural resources (Cabeza et al. 2018; Lövdén et al. 2010), which may lead to different outcomes of competition between degeneration and adaptation. These outcomes could be successful cognitive function maintenance for "healthy aging" people (Sachdev 2016) or the failure of adaptation with accumulated degeneration that may lead to brain disorders such as AD (Mesulam 1999). Although the overall evidence is in line with the important hypothesis that the brain adaptation and degeneration are highly related, few studies have provided direct empirical evidence and theoretical explanation at the brain network level.

Multimodal magnetic resonance imaging (MRI) technology and longitudinal experimental data collection provide effective and quantitative tools to investigate the brain network in this complex aging process. MRI provides structural information about brain morphology, which reflects the summation of cellular changes at the microscopic level (Sagi et al. 2012; Zatorre, Fields, and Johansen-Berg 2012). Functional MRI (fMRI), combined with graph theory, generates a set of functional network measures which help to perform novel analyses using modeling approaches (Bassett and Sporns 2017). Several data-driven models have been proposed for understanding relation between brain morphological changes (Douaud et al. 2014) and network dynamics (Cabral, Kringelbach, and Deco 2014; Shafto and Tyler 2014). Some studies have found the static correspondences between brain structural and functional connectivity (Honey et al. 2009; Honey, Thivierge, and Sporns 2010; Suárez et al. 2020). For example, structural damages such as white matter lesions or amyloid deposition are closely related to functional deficits (Bai et al. 2009; He et al. 2007; Johnston et al. 2008; Sperling et al. 2009). Recently, some studies have reported that the shortest topological pathway to "epicenters regions" can predict regional atrophy in patients of behavioral variant frontotemporal dementia and primary progressive aphasia (Brown et al. 2019; Mandelli et al. 2016). Those regions with shorter topological distance to "epicenters regions" have greater atrophy, implying that brain atrophy progress may rely on some specific brain functional connection patterns (Seeley 2017).

However, it remains essentially unknown until now whether and how the aging brain structure and function interact to achieve dynamical maintenance of the whole system. This possible dynamical interaction (Fig. 1a) between degeneration and adaptation determines the aging outcomes. Brain structural atrophy is a well-known and common phenomenon in the aging brain (Park and Reuter-Lorenz 2009, Fjell et al. 2014), however, functional network efficiency may be maintained in spite of this (Geerligs et al. 2015). To understand the borders between healthy aging and AD, it is also critical to investigate this complex interaction process across brain network during different types of aging processes (Nyberg et al. 2012). While above studies are mainly for single time stage, here we pay our main attention to the dynamic interaction at large-temporal scale across 3 testing stage of 6 years between brain structure and function during the aging process. Especially, we study how this coupling pattern achieves system resilience during healthy aging in a relatively long period. We find negative feedback interaction between brain structural atrophy and functional network efficiency, which could be explained by our theoretical model. Finally, we analyze how brain network resilience emerges among brain regions, to identify critical mitigation target.

**Results**

We use longitudinal multi-modal MRI data from the Sydney Memory and Aging Study (MAS). Structural MRI scans are obtained at baseline ($t_0$), 2-year ($t_1$) and 6-year follow-ups ($t_2$), and fMRI are also obtained at the latter two time points ($t_1$ and $t_2$). Totally, we analyze 63 participants and 303 scans (177 T1 and 126 fMRI) in this study. As shown in Fig. 1a, we can model the brain in structural and functional layers (Bassett and Sporns 2017; Bassett, Zurn, and Gold 2018), with total brain volume and network efficiency as structural and functional measures respectively (see Methods). We first ask how brain structure and function change separately in 6 years. While brain shows continuous atrophy in 6 years (Fig. 1b, Fig. S1), average global efficiency of brain function network keeps almost unchanged from $t_1$ to $t_2$ ($E_1 = 0.6368 \pm 0.1647$, $E_2 = 0.6472 \pm 0.1202$, Fig. 1c).

Through analysis of different time points, we find significant correlation between structural atrophy and functional efficiency of the brain network (Fig. 2). Specifically, there is a negative correlation between $Atrophy_{0,1} = (\frac{V_0 - V_1}{V_0})$ in the first stage and global efficiency $E_1$ (Fig. 2a), on the contrary $Atrophy_{1,2} = (\frac{V_1 - V_2}{V_1})$ in the next time stage is positively correlated with $E_1$ (Fig. 2b). In the first stage from $t_0$ to $t_1$, the larger the structural atrophy, the smaller the functional network efficiency becomes at $t_1$. Conversely, in the second stage from $t_1$ to $t_2$, the smaller the functional network efficiency at $t_1$, the smaller the structural atrophy becomes. The process of these two stages may suggest two critical behaviors for brain aging: degeneration and adaptation. In the first stage, brain declines in the structural layer due to atrophy, leading to the corresponding degeneration in the functional layer. In the second stage, the degenerated functional network adapts to slow down further structural atrophy. These two stages together represent one negative feedback mechanism, which we call the resilience function.

To demonstrate this resilience function, we divide these participants into two groups according to their structure-function interaction patterns. Then we standardized $Atrophy_{0,1}$, $E_1$ and $E_{2,1}$ with z-score respectively and measure the relative changes of each individual within each group (Fig. S2). The group with upwards convex trajectory (blue line) is defined as the Degeneration Group, and the other group showing the opposite pattern is defined as the Adaptation Group. We will show later that these 2 groups are possibly different period stages of one single resilience function. As shown in Fig. 3a, Degeneration Group starts with weak atrophy from $t_0$ to $t_1$ (compared with the whole population), and shows higher efficiency at $t_1$, which leads to higher atrophy from $t_1$ to $t_2$ in the next stage. It is suggested that high efficiency in brain function reflects an 'expensive' connection pattern, which might accelerate the spread of cumulative atrophy burden in the whole brain (Brown et al. 2019). The opposite process can be observed in the Adaptation Group, whose efficiency at $t_1$ is lower than the Degeneration Group. Because efficiency is significantly related with number of links and topology, lower global efficiency represents less links or wiring cost among brain regions, which in turn might help to slow the spread of brain atrophy. Our findings here represent empirical evidence for the degeneration-adaptation antagonism interaction.

Surprisingly, when comparing functional efficiency between these two groups (Fig. 3b), we find that after four-year changes from $t_1$ to $t_2$, their efficiency distribution converges to similar mode: efficiency of Degeneration Group decreases from $t_1$ to $t_2$, while efficiency of Adaptation Group increases from $t_1$ to $t_2$. This suggests the existence of one resilience function maintaining the brain network operation during the aging process. Resilience is usually defined as the system's ability to absorb the perturbation and recover to the original functional state (Gao, Barzel, and Barabási 2016). Here it seems that brain networks can absorb the aging perturbation and spontaneously recover to its efficiency equilibrium. After distinguishing the two groups, we re-examine the relationship between $E_1$ and $Atrophy_{1,2}$ (Fig. 3c) for each group separately. We find that the significant efficiency-atrophy correlation in the total group is mainly caused by the Degeneration Group. This suggests dedicated transition mechanism that there exists a critical point for network efficiency $E$, above which the positive interdependence between network function and structure atrophy starts. For the Adaptation Group, brain structure atrophy is relatively small and stable because their efficiency is below the threshold.

To understand the resilience function during aging, we construct theoretical model. Fig. 4a demonstrates the basic idea of our model that the negative feedback mechanism keeps the efficiency and atrophy speed near the equilibrium value for health aging. When efficiency of brain functional network is higher than the equilibrium value, atrophy will become faster because the functional network requires more energy, leading to larger oxidative stress that accelerate atrophy. Faster atrophy leads to more neural link losses. As the repair ability of network cannot follow the speed of damage, efficiency decreases. When the efficiency is lower than the equilibrium value, atrophy becomes slower with smaller oxidative stress. When the speed of atrophy is slow enough, the repair ability of the network can follow the speed of damage, and the efficiency can bounce back again through self-organization.

To explain our main idea, we model the dynamics of efficiency and atrophy via the following discrete-time equations:

$$A(t) = k_1 E(t-1) + C_A Z_1(t), \qquad (1)$$

$$E(t) - E(t-1) = -k_e E(t-1) - k_2 A(t) + C_E Z_2(t), \qquad (2)$$

where $A(t) = \frac{V(t-1)-V(t)}{V(t-1)}$ denotes atrophy from step $t-1$ to step $t$, and $E(t)$ denotes efficiency at step $t$. Note that $A(t)$ and $E(t)$ have been normalized (see SI-Theoretical model) so that their equilibrium value are zero. In Eq. (1) and Eq. (2), $k_1 > 0$ represents the effect that high efficiency causes fast atrophy, $k_2 > 0$ represents the effect that fast atrophy makes efficiency decreasing, $k_e$ represents the negative feedback of efficiency rather than atrophy, $C_A > 0$ and $C_E > 0$ represents the noise in dynamics. Here $Z_1(t)$ and $Z_2(t)$ at different time are assumed independent standard normal random variables.

To illustrate that this model can capture the resilience mechanism shown in Fig. 4a, we simulate brains aging by our model as shown in Fig. 4b-4e. In Fig. 4b, we display the volume (the initial value of volume is set to 1) and efficiency data generated by the model over 20 years. While volume is decreasing with time, efficiency is fluctuating around its equilibrium. The first arrow in Fig. 4b shows that slowing atrophy will make efficiency larger, which in turn will accelerate the atrophy rate in the next step shown by the second arrow. We can use this model to reproduce two correlation relationships consistent with the observed results above. As shown in Fig. 4c and Fig. 4d, theoretical correlation between efficiency and atrophy shows similar pattern to our empirical findings in Fig. 2a and Fig. 3a. We further analyze theoretically how the correlation changes with all parameters (see SI-Theoretical model Section S2). For analyzing how brain evolves in different states, states are divided into four kinds of states:

State 1: high efficiency ($E(t) > 0$) and fast atrophy ($\frac{A(t)+A(t+1)}{2} > 0$).

State 2: high efficiency ($E(t) > 0$) and slow atrophy ($\frac{A(t)+A(t+1)}{2} < 0$).

State 3: low efficiency ($E(t) < 0$) and slow atrophy ($\frac{A(t)+A(t+1)}{2} < 0$).

State 4: low efficiency ($E(t) < 0$) and fast atrophy ($\frac{A(t)+A(t+1)}{2} > 0$).

Under randomly sampled parameters, we estimate the average probability $p_{ij}$ of moving from state $i$ to state $j$ (see SI-Theoretical model Section S3). We find that the mode shown in Fig. 4e is an universal mode for the interaction between atrophy and efficiency, which can be observed from most of parameter configurations.

With a clear picture of aging resilience in the macroscopic scale, here we analyze in the microscopic scale how the brain functional network evolves during different periods. K-core decomposition breaks down the network into 'cores' by recursively pruning the least connected nodes (Alvarez-Hamelin et al. 2005). The k-core decomposition allows easy comparison of structural properties within different networks on a hierarchical backdrop. As shown in Fig. 5a, from $t_1$ to $t_2$, the Degeneration Group breaks faster with successive removal of node shells. The inverse process has been observed for the Adaptation Group. A possible explanation is that during the decomposition process at $t_1$, the network of Degeneration Group seems more centralized, while the network of Adaptation Group is easier to break into multiple centers. For the next time period, we see the opposite changes in the network structure. At $t_2$, the networks of the Degeneration Group become more decentralized, while the networks of the Adaptation Group become more hierarchically connected. The network structure plotted during k-core decomposition process can be seen in Fig. S4 and Fig. S5. This finding confirms our previous observation that these two groups would converge to similar functional equilibrium under one resilience function, where two networks become much similar under hierarchical backdrop at $t_2$. To further explore hierarchical structure of the network, we perform a rich club analysis. The 'rich-club' phenomenon refers to the tendency of nodes with high centrality, the dominant elements of the system, to form tightly interconnected communities (Colizza et al. 2006). As shown in Fig. 5b and Fig. 5c, for the Degeneration Group, the functional network changed from high to low enrichment, accompanied by rich club connectivity transition from high-degree to medium-degree. A reasonable supposition is that connections between hubs are shrinking due to fast atrophy. For the Adaptation Group, the inverse process has been observed with the formation of centralized region with dense connection between hubs.

During the aging process, different brain regions play different roles. To identify the key region in the resilience function, we separate all cortical areas into 7 modules according to the Yeo template (Thomas Yeo et al. 2011), yielding 28 sub-networks within and between brain regions (Fig. 6a). We can observe that the efficiency decrease in different brain regions is homogeneous, while efficiency increase is heterogeneous. From period $t_1$ to $t_2$ for Adapation Group, the top regions with increased efficiency are all limbic-related, including Visual-Limbic, Somatomotor-Limbic, Dorsal Attention-Limbic, Frontoparietal-Limbic and Limbic. In the process of redistributing efficiency, Limbic regions may be the dominant area because of its higher neuroplasticity (Mesulam 1999). In order to compare the difference in efficiency redistribution between the two groups, we calculated the normalized efficiency redistribution difference (Fig. 6b). It can be seen that all areas related to Limbic regions have large redistribution differences. This indicates that the Limbic area may be a possible driving factor for the resilience function, which calls for further experimental research.

**Discussion**

From the complex system view, we conduct an analysis of brain aging based on 303 multimodal MRI scans with 3-time-point longitudinal data from non-dementia participants. Through empirical exploration, we reveal resilience function as the negative feedback interaction between brain structural atrophy and functional network efficiency, where the brain alternately shows losses and compensation. We also develop theoretical model for the resilience function reflected in the dynamical balance between the continuous wear and tear of the brain and self-compensation during the aging process. The greater the network efficiency, the higher the running costs to the brain (Bullmore and Sporns 2012). This may put higher burden on the brain hardware and accelerate the brain atrophy. At this point, the degeneration rate of the brain is relatively greater than the adaptation of self-compensation. Later, brain atrophy will damage functional network wiring, which in turn reduces network efficiency. As network efficiency of the brain turns down to lower level, and the running cost of the brain becomes relatively reduced. Although the brain atrophy caused by aging at this period is still affecting brain function, the compensation mechanism of the brain will outperform the aging losses. New network connection patterns are generated through the reorganization to recover the normal function of the brain.

Our research provides understanding of brain aging from the viewpoint of system resilience. Resilience is critical dynamic process including absorption, adaptation and recovery, which has been found in natural and man-made systems (Adger et al. 2005; Gao et al. 2016; Ponce-Campos et al. 2013; Southwick and Charney 2012; Zhang et al. 2019). We argue that degeneration and adaptation are two interactive processes of one resilience function which helps to maintain dynamically risk-balanced successful aging. More importantly, there are usually early-warning signals for the resilience process. For example, the recovery rate to the equilibrium point would become slow when approaching the critical point of transition (Scheffer et al. 2012). Based on our findings, it would be urgent and critical in the next step to examine whether there exists such critical point of transition between healthy aging and disease, based on which new health aging management methods could be designed.

We find that Limbic brain regions are highly active in the process of rebuilding or rupturing edges with other brain regions. The Limbic brain area has been identified as area with high adaptation (Mesulam 1999), which includes 3 distinct overlapping networks of memory, emotion and introspection (Catani, Dell'Acqua, and Thiebaut de Schotten 2013). Elders show a diminished use of temporo-Limbic regions in facial emotion processing (Gunning-Dixon et al. 2003). Negative emotion such as stress may accelerate frontoamygdala development, which may offer a protection against accelerated biological aging (Miller et al. 2021). These studies show that Limbic regions are closely related to aging with emotion state. Alzheimer's pathological transmission studies show that neuro-degeneration generally start from the Limbic system such as the entorhinal cortex and hippocampus, pass through the posteromedial cortex and the inferior parietal lobules, and finally spread to the prefrontal cortex (Arnold et al. 1991; Kapogiannis and Mattson 2011). Combined with our findings, these characteristics show that the Limbic brain region, with high vulnerability and high adaptation at the same time, may play a fundamental role in the resilience function of human brain aging regulation process.

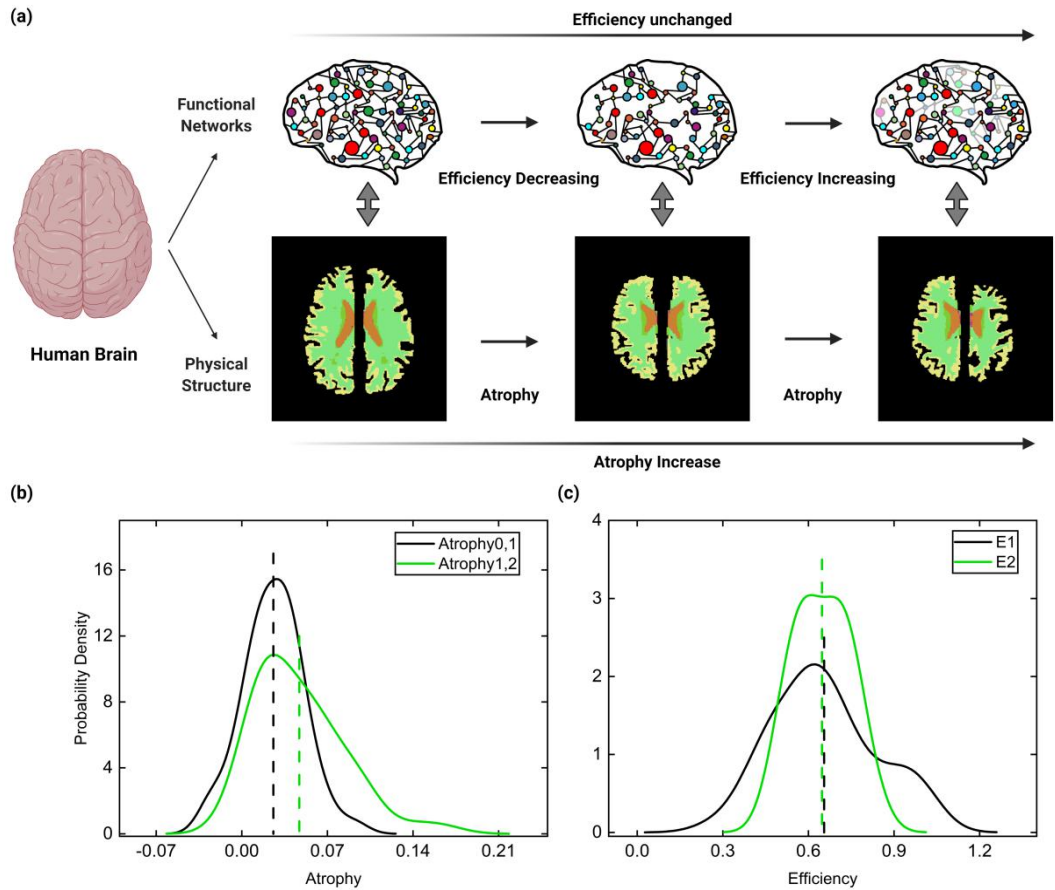

**Figure 1**. The differential trajectories of brain structure and function. (a) The global efficiency of the network decreases and increases, forming a resilience mode, with continuous atrophy with different speed. (b) The histogram of brain atrophy for two groups. (c) The histogram of global efficiency for two groups.

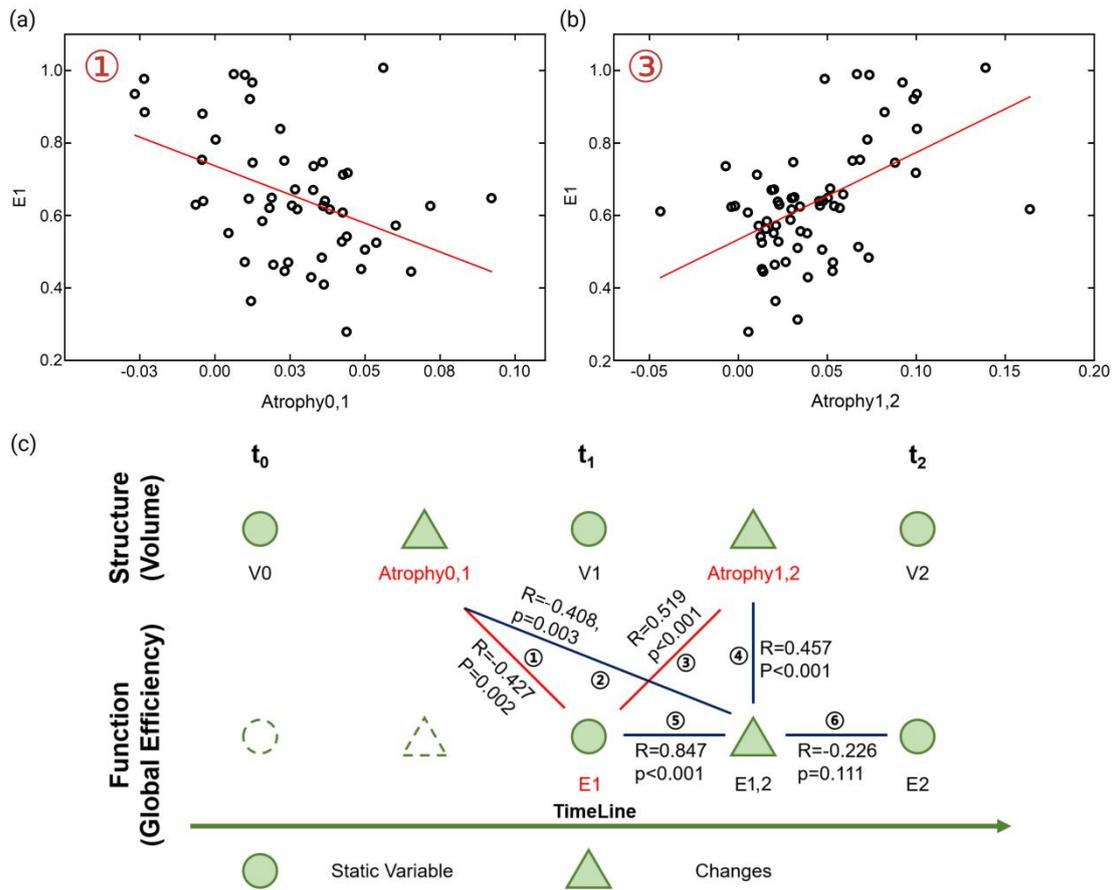

**Figure 2**. The correlation between brain structure and function during aging. (a) and (b) show the correlation of brain global efficiency at $t_1$ ($E_1$) and brain atrophy: (a) before $t_1$ and (b) after $t_1$. (c) The upper row represents the structural variables changes, and the lower row shows the functional variables. The circles represent the values at that time point, and the triangles represent the changes between two time points. The functional data have only two time points due to lack of data at the baseline time. The lines and circled numbers show the significant correlated pairs with p<0.05.

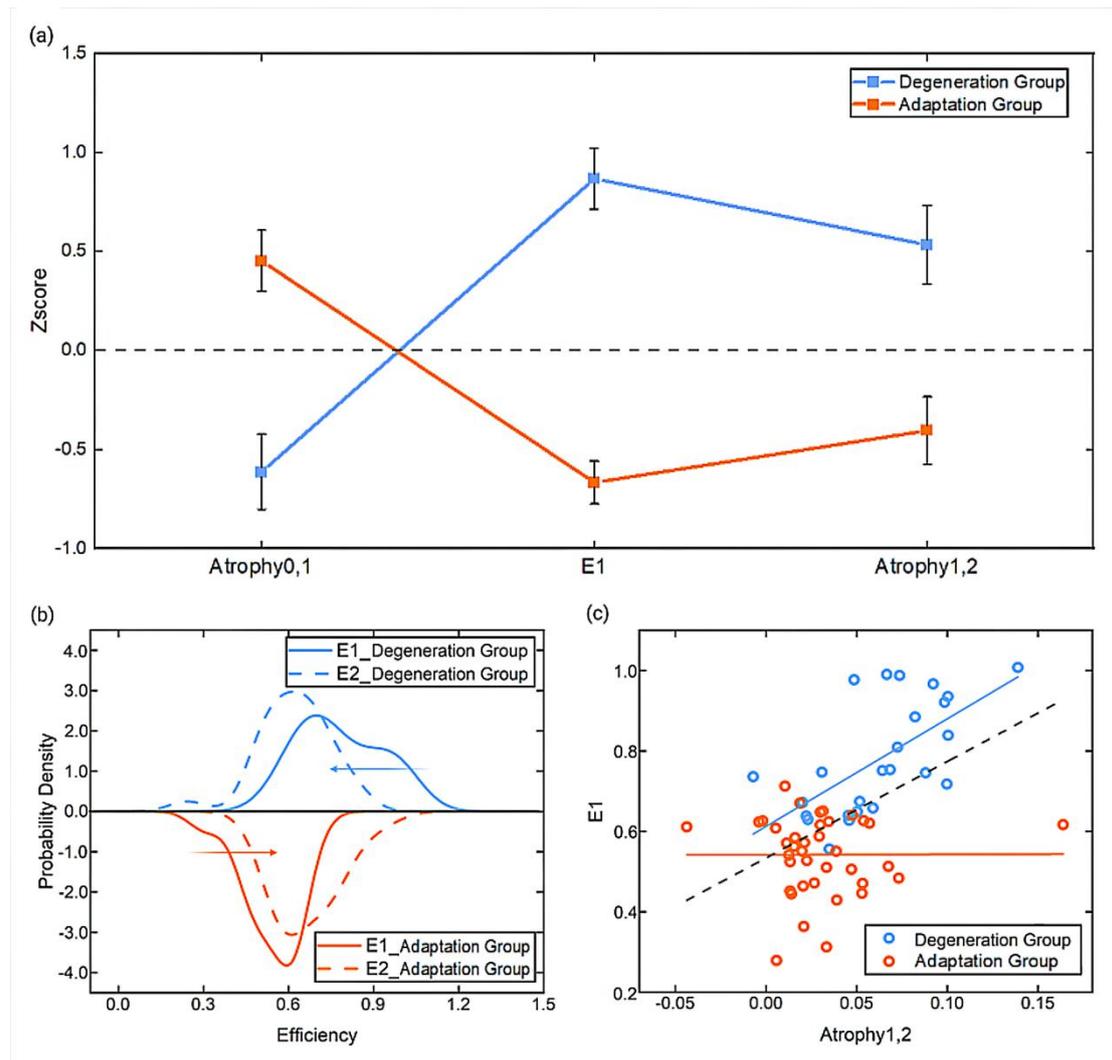

**Figure 3**. Degeneration and Adaptation. (a) The trajectories of "atrophy-efficiency-atrophy" for the Degeneration Group (blue) and the Adaptation Group (orange). For each time point, we calculate z-score for the whole participants respectively. Then we divide the whole participants into 2 groups according to the shape of trajectories. (b) Probability density function of the efficiency for both groups at two time points. (c) Distinction between the correlation of efficiency and atrophy for the two groups.

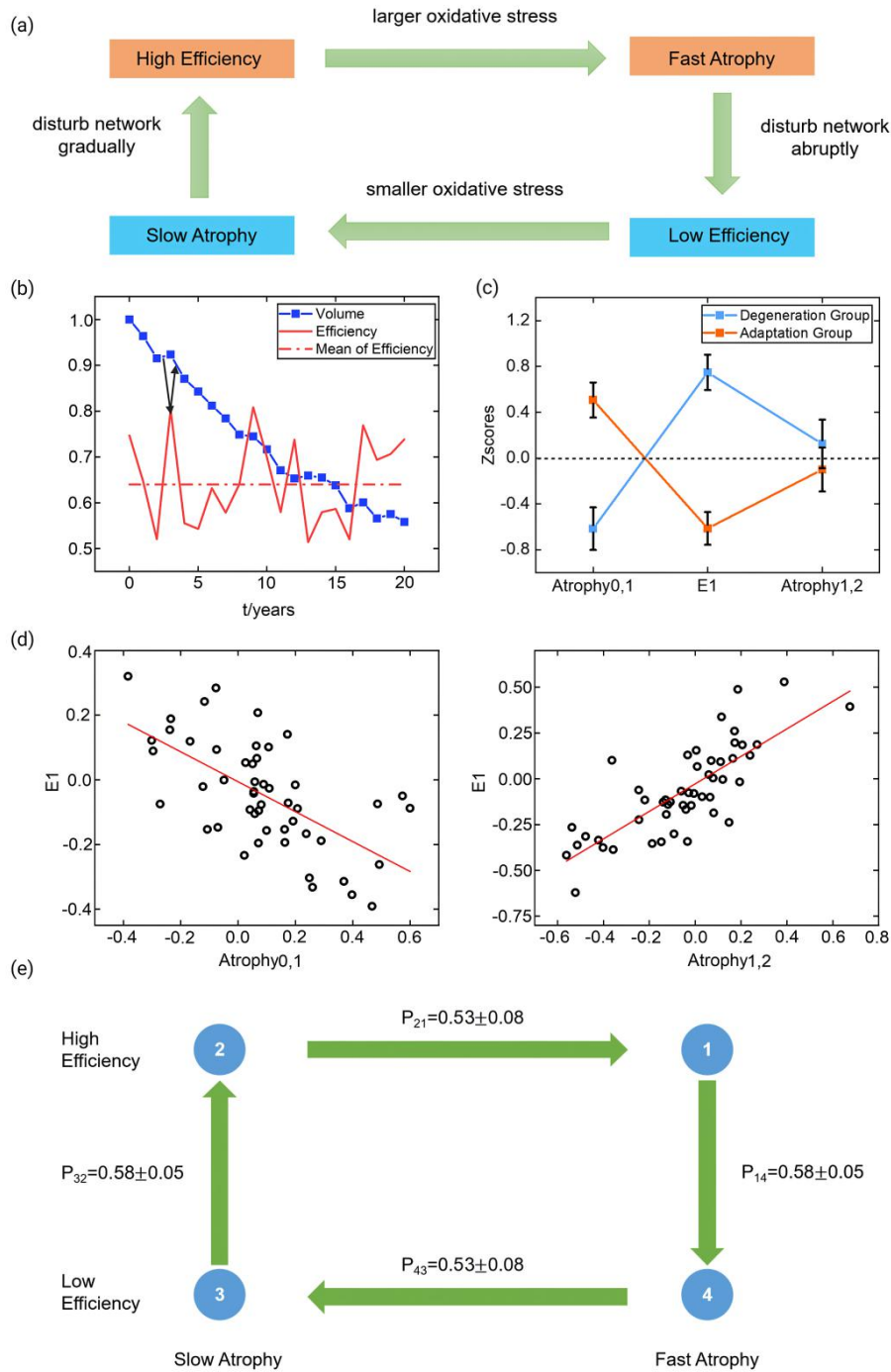

**Figure 4.** Theoretical model explains the resilience mode how efficiency and atrophy speed oscillate with each other. (a) The negative feedback of efficiency and atrophy. (b) Volume and efficiency generated by model, and the initial value of volume is 1. (c) Model generates the trajectories of "atrophy-efficiency-atrophy" for the Degeneration Group (blue) and the Adaptation Group (orange). (d) Correlation between atrophy and efficiency from theoretical model. (e) The average transition probability for 10000 sets of parameters. Only $P_{ij}$ above 0.27 are shown.

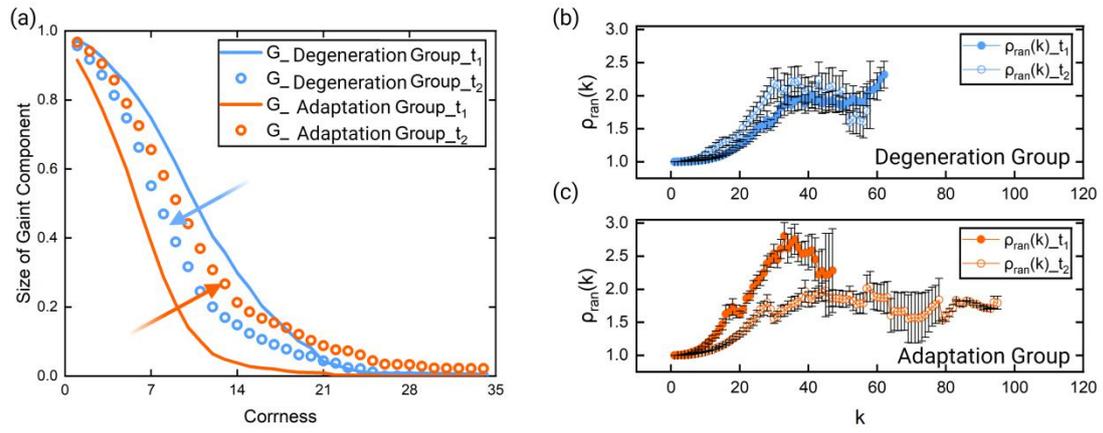

**Figure 5**. Network resilience during aging. (a) K-core decomposition for Degeneration Group and Adaptation Group at $t_1$ and $t_2$ respectively. (b) 'Rich club' analysis for two groups at two time points.

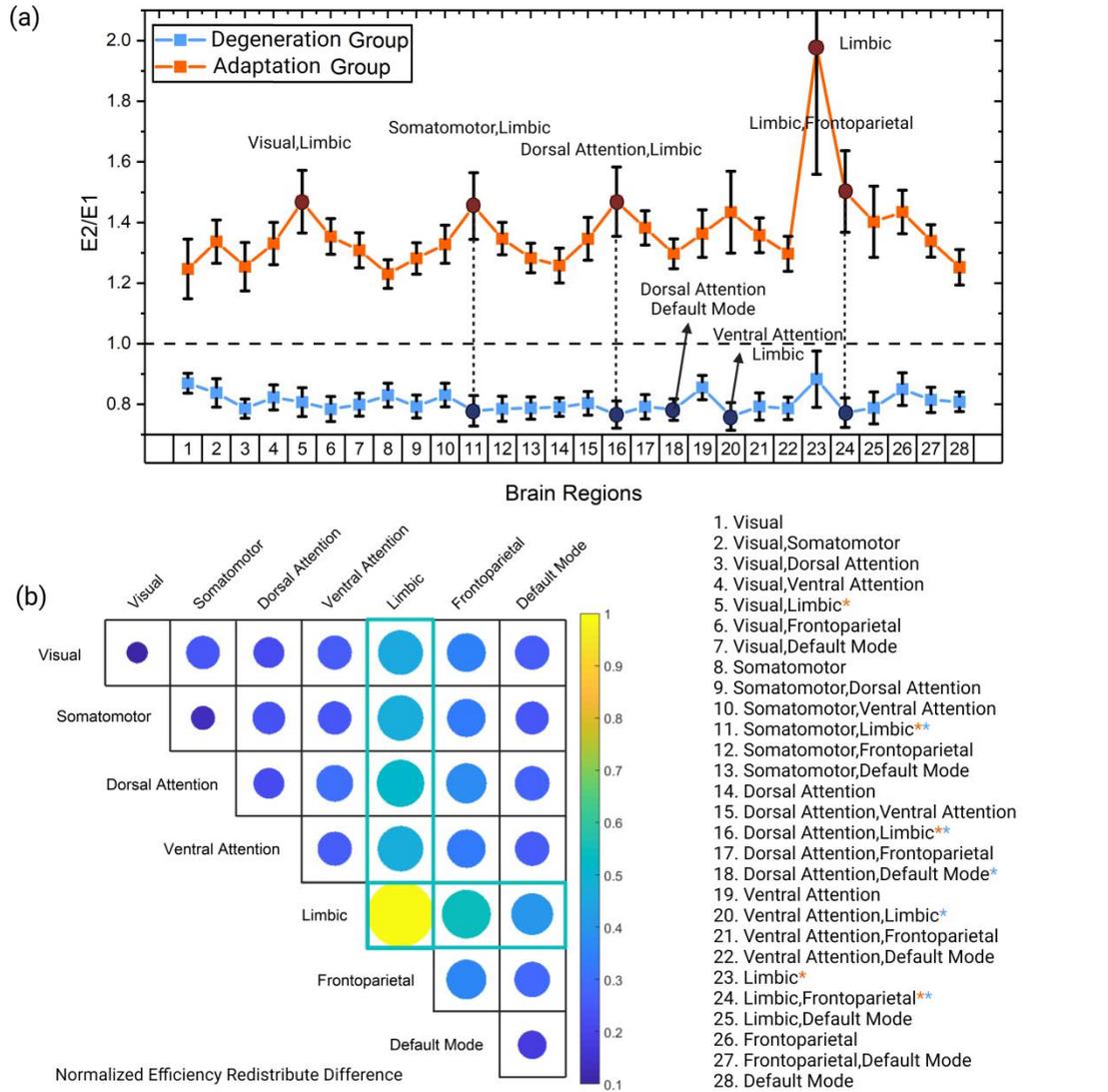

**Figure 6**. Critical role of Limbic region. (a) The brain is divided into 7 brain modules (Yeo template), with 7 internal brain modules and 21 cross brain modules. For the Degeneration Group, $E_2/E_1$ are less than 1. For Adaptation Group, $E_2/E_1$ are larger than 1. The top 5 largest $E_2/E_1$ among 28 brain modules and five brain regions with smallest $E_2/E_1$ are highlighted with black circles. (b) The normalized efficiency redistributes differently among these modules. The values of $E_2/E_1$ in different brain modules between the two groups are compared. Limbic region and regions associated with Limbic show great difference between two groups.